\documentstyle[preprint,aps]{revtex}
\tightenlines
\title {Controlled light storage in a double lambda
system}
\author{A. Raczy\'nski
\footnote{email: raczyn@phys.uni.torun.pl} and J. Zaremba}
\address{Instytut Fizyki, Uniwersytet Miko\l aja Kopernika,
       ul.Grudzi\c{a}dzka 5,
       87-100 Toru\'n, Poland }
\begin{document}

\maketitle
\begin{abstract}
It is shown theoretically that after light storing in a medium of
four-level atoms it is possible to release a new pulse of a
different frequency, the process being steered by another driving
beam. It is also possible to store one pulse and to release two
different ones, with their time separation and heights being
controlled.
\end{abstract}
\newpage
It has been shown both theoretically  and experimentally (see,
e.g., Refs \cite{t1,t2,t3,e1,e2} and a review \cite{n1}) that a
light pulse, propagating in a medium composed of three-level atoms
in a $\Lambda$ configuration, suitably driven by another pulse,
can be stopped and later released in a controlled way. The process
is interpreted in terms of inducing a transient Raman coherence
between two lower atomic states or, in the language of
quasi-particles, in terms of an adiabatic evolution of the
so-called dark-state polariton \cite{t1,t2}. Natural questions
arise whether it is possible to convert the stored light in a
controlled way into a pulse of a frequency different from that of
the stopped one, or into more pulses of different frequencies, or
what happens if the Raman coherence is due to a transient
absorption of more than one pulse. Some experimental results
concerning light frequency conversion during the process of its
storing and retrieving in a four-level system have recently been
mentioned by Matsko {\em et al.} \cite{n1}. In this paper we
present a complete theoretical analysis of the situations when a
four-level-atom medium is driven by two control laser fields and,
after stopping one or two probe pulses, one or two pulses can be
released. We also point out the possibility of controlling the
process by a relative time shift of the control fields.

Consider a quasi one-dimensional medium of four-level atoms with
two lower metastable states $|b>$ and $|c>$ and two upper states
$|a>$ and $|d>$ (Fig.\ref{fig1}). The position of an atom is
described by the variable $z$, which is considered continuous. The
weak signal fields 1 and 3 couple the initial state $|b>$ with
$|a>$ and $|d>$, while the control fields 2 and 4 couple $|c>$
with $|a>$ and $|d>$, respectively. The interaction Hamiltonian is
$V=-\hat{d}\sum_{j=1}^{4} \epsilon_{j}\cos\phi_{j}$, with
$\phi_{j}=\omega_{j} t-k_{j} z$, $\epsilon_{j}=\epsilon_{j}(z,t)$
being slowly varying envelopes and all the fields having the same
linear polarization. The matrix elements of the dipole moment
$d_{1}=(\hat{d})_{ab}$, $d_{2}=(\hat{d})_{ac}$,
$d_{3}=(\hat{d})_{bd}$, $d_{4}=(\hat{d})_{cd}$ are taken real.
Resonant conditions concerning all the couplings are assumed, i.e.
$\hbar \omega_{1}=E_{a}-E_{b}$, $\hbar \omega_{2}=E_{a}-E_{c}$,
$\hbar \omega_{3}=E_{d}-E_{b}$, $\hbar \omega_{4}=E_{d}-E_{c}$.

The evolution equation $i\hbar \dot{\rho}=[H,\rho]$ for the
density matrix $\rho=\rho(z,t)$ for an atom at position $z$, after
making the rotating-wave approximation, transforming-off the
rapidly oscillating factors: $\rho_{ab}=\sigma_{ab}
\exp(-i\phi_{1})$, $\rho_{ac}=\sigma_{ac} \exp(-i\phi_{2})$,
$\rho_{bc}=\sigma_{bc} \exp[i(\phi_{1}-\phi_{2})]$,
$\rho_{db}=\sigma_{db} \exp(-i\phi_{3})$, $\rho_{dc}=\sigma_{dc}
\exp(-i\phi_{4})$, $\rho_{ad}=\sigma_{ad}
\exp[i(\phi_{3}-\phi_{1})]$, $\rho_{ii}=\sigma_{ii}$, and after
adding relaxation terms describing the spontaneous emission within
the system, takes the form
\begin{eqnarray}
 i \hbar \dot{\sigma}_{aa}&=&
-\frac{1}{2}\epsilon_{1}d_{1}(\sigma_{ba}-\sigma_{ab})-
\frac{1}{2}\epsilon_{2}d_{2}(\sigma_{ca}-\sigma_{ac})-i \hbar
(\Gamma^{a}_{b}+\Gamma^{a}_{c}) \sigma_{aa},\nonumber\\
 i \hbar\dot{\sigma}_{bb}&=&
 -\frac{1}{2}\epsilon_{1}d_{1}(\sigma_{ab}-\sigma_{ba})-
\frac{1}{2}\epsilon_{3}d_{3}(\sigma_{db}-\sigma_{bd})+i \hbar
\Gamma^{a}_{b} \sigma_{aa} +i \hbar\Gamma^{d}_{b}
\sigma_{dd},\nonumber\\
 i \hbar \dot{\sigma}_{cc}&=&
-\frac{1}{2}\epsilon_{2}d_{2}(\sigma_{ac}-\sigma_{ca})-
\frac{1}{2}\epsilon_{4}d_{4}(\sigma_{dc}-\sigma_{cd})+i \hbar
\Gamma^{a}_{c} \sigma_{aa} +i \hbar\Gamma^{d}_{c}
\sigma_{dd},\nonumber\\
 i \hbar\dot{\sigma}_{dd}&=&
-\frac{1}{2}\epsilon_{3}d_{3}(\sigma_{bd}-\sigma_{db})-
\frac{1}{2}\epsilon_{4}d_{4}(\sigma_{cd}-\sigma_{dc})-i \hbar
(\Gamma^{d}_{b} +\Gamma^{d}_{c}) \sigma_{dd},\nonumber\\
 i \hbar\dot{\sigma}_{ab}&=&-\frac{1}{2}\epsilon_{1} d_{1} (\sigma_{bb}-
 \sigma_{aa})-\frac{1}{2}\epsilon_{2}d_{2}\sigma_{cb}
 + \frac{1}{2}\epsilon_{3}d_{3}\sigma_{ad}-
 \frac{i \hbar}{2}(\Gamma^{a}_{b}+\Gamma^{a}_{c})\sigma_{ab},\\
 i \hbar\dot{\sigma}_{ac}&=&-\frac{1}{2}\epsilon_{1} d_{1} \sigma_{bc}-
\frac{1}{2}\epsilon_{2}d_{2}(\sigma_{cc}-\sigma_{aa})
 + \frac{1}{2}\epsilon_{4}d_{4}\sigma_{ad}-
 \frac{i\hbar}{2}(\Gamma^{a}_{b}+\Gamma^{a}_{c})\sigma_{ac},\nonumber\\
 i \hbar \dot{\sigma}_{ad}&=&-\frac{1}{2}\epsilon_{1} d_{1} \sigma_{bd}-
\frac{1}{2}\epsilon_{2}d_{2}\sigma_{cd}+
\frac{1}{2}\epsilon_{3}d_{3}\sigma_{ab}
 + \frac{1}{2}\epsilon_{4}d_{4}\sigma_{ac}-
 \frac{i\hbar}{2}(\Gamma^{a}_{b}+\Gamma^{a}_{c}+\Gamma^{d}_{b}+\Gamma^{d}_{c})
 \sigma_{ad},\nonumber\\
 i \hbar \dot{\sigma}_{bc}&=&-\frac{1}{2}\epsilon_{1} d_{1} \sigma_{ac}+
\frac{1}{2}\epsilon_{2}d_{2}\sigma_{ba}-
\frac{1}{2}\epsilon_{3}d_{3}\sigma_{dc}
 + \frac{1}{2}\epsilon_{4}d_{4}\sigma_{bd},\nonumber\\
 i \hbar \dot{\sigma}_{bd}&=&
-\frac{1}{2}\epsilon_{1}d_{1}\sigma_{ad}
-\frac{1}{2}\epsilon_{3}d_{3}(\sigma_{dd}-\sigma_{bb})+
\frac{1}{2}\epsilon_{4}d_{4}\sigma_{bc}
-\frac{i\hbar}{2}(\Gamma^{d}_{b}+\Gamma^{d}_{c})\sigma_{bd},\nonumber\\
 i \hbar\dot{\sigma}_{cd}&=&
-\frac{1}{2}\epsilon_{2}d_{2}\sigma_{ad}
+\frac{1}{2}\epsilon_{3}d_{3}\sigma_{cb}-
\frac{1}{2}\epsilon_{4}d_{4}(\sigma_{dd}-\sigma_{cc})
-\frac{i\hbar}{2}(\Gamma^{d}_{b}+\Gamma^{d}_{c})\sigma_{cd},\nonumber
\end{eqnarray}
where $\Gamma^{a}_{b}$ is the decay rate of the state $|a>$ to
$|b>$ etc.

The propagation equations for the signal fields 1 and 3 are
written as usual in the slowly varying envelope approximation
\cite{s1} after rejecting the second space and time derivatives of
$\epsilon_{j}$. In the conditions of the resonance they read
\begin{eqnarray}
\frac{\partial \epsilon_{1}}{\partial z}+\frac{1}{c}
\frac{\partial \epsilon_{1}}{\partial t}=i N
d_{1}\frac{\omega_{1}}{\epsilon_{0}c} \sigma_{ab},\nonumber\\
\frac{\partial \epsilon_{3}}{\partial z}+\frac{1}{c}
\frac{\partial \epsilon_{3}}{\partial t}=i N
d_{3}\frac{\omega_{3}}{\epsilon_{0}c} \sigma_{db},
\end{eqnarray}
where $N$ is the atom density, $\epsilon_{0}$ is the vacuum
electric permittivity and use has been made of the fact that in
the resonance conditions $\sigma_{ab}$ and $\sigma_{db}$ are
imaginary numbers. Similarly as in earlier papers, we have
neglected propagation effects for the driving fields, i.e.
$\epsilon_{2,4}=\epsilon_{2,4}(t)$.

Eqs (1) and (2) have been solved numerically in the moving window
frame of reference: $t'=t-z/c,z'=z$ using the method described by
Shore \cite{s2}. Switching the driving fields on and/or off was
modeled by a hyperbolic tangent. The initial probe pulse was taken
as the sine square shape $\epsilon_{1}(0,t)=\epsilon_{10}
\sin^{2}[\pi(t-t_{2})/(t_{2}-t_{1})]\Theta(t-t_{1})\Theta(t_{2}-t)$,
while the initial condition for the atomic part was
$\sigma_{bb}(z,0)=1$, with other matrix elements equal to zero.

We have performed simulations for somewhat arbitrarily chosen
data, being however of realistic orders of magnitude. The model
atomic energies were $E_{a}=-0.10$ a.u., $E_{b}=-0.20$ a.u.,
$E_{c}=-0.18$ a.u., $E_{d}=-0.05$ a.u., all the relaxation rates
due to the spontaneous emission  were taken
$\Gamma^{a,d}_{b,c}=2.4\times 10^{-9}$ a.u., from which the
corresponding transition dipole moments were calculated.  The
length of the atomic sample was $3\times 10^{7}$ a.u. (1.6 mm) and
its density $3\times 10^{-13} $ a.u. $(2\times 10^{12}$
cm$^{-3}$). The initial signal pulse length was $10^{11}$ a.u.
(2.4 $\mu$s) and $\epsilon_{10}=10^{-10}$ a.u. (which corresponds
to the power density of 3.5$\times10^{-4}$ Wcm$^{-2}$); the
maximum value of the control field amplitudes was $1.2
\times10^{-9}$ a.u. (50 mWcm$^{-2}$).

The upper part of Fig. \ref{fig2} shows for comparison the results
obtained by solving Eqs (1) and (2) in the case corresponding to
the recently studied light storage in a single $\Lambda$ system
(cf. the results of Ref. \cite{e1}). The pulse's electric field is
shown as a function of the "local" time $t'$ (which for our data
is almost equal to $t$). The left peak of the signal pulse is the
untrapped fraction of the incoming pulse, i.e. its fraction
transmitted by the medium before the control pulse has been
switched off. The rest of the pulse is trapped and later released
(the right peak) due to turning the control pulse back on. We have
checked that, as expected, the time interval between the left and
right peaks is the same as that between the instants of switching
the control pulse off and on.

The lower part of Fig. \ref{fig2} shows the new result predicted
in this paper. As before, the signal pulse $\epsilon_{1}$ is
trapped due to switching the control field $\epsilon_{2}$ off and
its untrapped part is observed. However, in this case, after some
time the {\em other} control field $\epsilon_{4}$ is turned on,
which results in generating a {\em new} pulse $\epsilon_{3}$ of
frequency different from that of the original trapped signal
pulse. Also in this situation the instant of appearing of the
released pulse (now the field 3) is controlled by choosing the
moment of the switch-on of the driving field (now the field 4).
Numerical calculations show that the height and shape of this new
released pulse strongly depend on the process parameters, in
particular on the maximum value and slope of the control field 4.

From our observations it follows that it should be possible to
trap a single pulse and then to release two or more pulses. Such a
possibility is demonstrated in Figs 3a and 3b, which show the
released pulses in the cases in which the driving field
$\epsilon_{4}$ was switched on, respectively, before and after the
driving field $\epsilon_{2}$. The difference of the pulses heights
shows that by choosing the order of the turn-on of the driving
pulses and their time delay we can influence the fraction of the
atomic coherence taken over by each of the released pulses. We
have also obtained the results (not shown here) which prove that
it is possible to simultaneously store two pulses and release one
or two new ones.

Extrapolating the interpretation concerning a single $\Lambda$
system one may look, analogously as in Ref. \cite{t1}, for the
solutions of Eqs (1) and (2), assuming a perturbative, adiabatic
and relaxationless evolution. It follows then that in particular

\begin{equation}\sigma_{bc}=-\frac{\epsilon_{1}d_{1}}{\epsilon_{2}
d_{2}}=-\frac{\epsilon_{3}d_{3}}{\epsilon_{4}d_{4}}.
\end{equation}

Those approximations allow one to find a shape-preserving solution
of the corresponding Maxwell-Bloch equations
$\Psi(z,t)=\Psi(z-\int_{0}^{t} v(t')dt',t=0)$, $\Psi$ being a
combination of field and atomic variables
\begin{equation}
\Psi=\frac{\frac{d_{2}\epsilon_{2}}{d_{1}}
\epsilon_{1}-\frac{2N\hbar\omega_{1}}{\epsilon_{0}}\sigma_{bc}+
\frac{d_{4}\epsilon_{4}\omega_{1}}{d_{3}\omega_{3}}\epsilon_{3}}
{\sqrt{\frac{d_{2}^{2}}{d_{1}^{2}}\epsilon_{2}^{2}+
\frac{2N\hbar\omega_{1}}{\epsilon_{0}}+\frac{d_{4}^{2}
\omega_{1}}{d_{3}^{2}\omega_{3}}\epsilon_{4}^{2}}}.
\end{equation}
The velocity $v$ is given by
\begin{equation}
v=c \frac{\frac{d_{2}^{2}}{d_{1}^{2}}\epsilon_{2}^{2}+
+\frac{d_{4}^{2} \omega_{1}}{d_{3}^{2}\omega_{3}}\epsilon_{4}^{2}}
{\frac{d_{2}^{2}}{d_{1}^{2}}\epsilon_{2}^{2}+
\frac{2N\hbar\omega_{1}}{\epsilon_{0}}+\frac{d_{4}^{2}
\omega_{1}}{d_{3}^{2}\omega_{3}}\epsilon_{4}^{2}}.
\end{equation}

However, after switching the field 4 on, the solution $\Psi$ tends
to $\sqrt{\frac{\omega_{1}}{\omega_{3}}}\epsilon_{3}$ instead of
$\epsilon_{3}$, which means that this solution cannot fully
characterize the pulse release phase of the evolution. The latter
would be described by an adiabatic evolution of
$\Psi'=\sqrt{\frac{\omega_{3}}{\omega_{1}}} \Psi$. This means that
light storing and releasing in a double $\Lambda$ system must
include a nonadiabatic phase; the measure of nonadiabaticity is
the ratio of the frequencies of the two signals. This has been
checked numerically: we have compared the computed coherence
$\sigma_{bc}(t')$ with its adiabatic approximations given by both
parts of Eq. (3). While the evolution of $\sigma_{bc}$ in the
storage phase was well reproduced by the part of Eq. (3) including
the fields 1 and 2, this was not the case for the part including
the fields 3 and 4 in the release phase. The coherence calculated
from the two parts of Eq. (3) exhibited a discontinuity, which
could be reduced by artificially introducing the factor
$\sqrt{\frac{\omega_{3}}{\omega_{1}}}$. The failure of the
adiabatic approximation in this case is connected with the fact
that a four level atom in the conditions of resonance, dressed by
the four interactions (1-4) satisfying the second equality in Eq.
(3), has a double real eigenvalue (equal to the bare energies).
Thus the assumptions of the adiabatic theorem are not satisfied.
Considerations analogous to those presented above but performed in
the formalism of second quantization would allow an interpretation
of the process in terms of quasiparticles. As in the case of a
single $\Lambda$ both the storage and release phases can be seen
as an adiabatic evolution of a dark state polaritons, but here the
two polaritons in the two phases are not identical and a
nonadiabatic transformation of one into another must take place.

The above results suggest new interesting possibilities of
controlling light propagation effects. A systematic quantitative
analysis of the dependence of the pulse shapes, as well as of the
time evolution of the atomic properties, on the numerous
accessible control parameters will be the subject of a future
work.

\newpage
\noindent

\begin{figure}
\caption{The double $\Lambda$ scheme of levels and couplings; the
indices 1 and 3 refer to signal fields and 2 and 4 - to control
fields.} \label{fig1}
\end{figure}

\begin{figure}
\caption{The shape of the signal pulses at the end of the sample
as the function of the "local" time $t'$; also shown are the
control pulses (for pictorial reasons their values were reduced by
the factor of 0.03); (a) the case of a single $\Lambda$ system:
$\epsilon_{3,4}\equiv 0$, $\epsilon_{1}$, solid line;
$\epsilon_{2}$, dashed line; for comparison is also shown the
shape of initial pulse, dotted line; (b) the case of a double
${\Lambda}$ system: the transmitted part of the original signal
pulse $\epsilon_{1}$, solid line; the newly created signal pulse
$\epsilon_{3}$, dashed line; $\epsilon_{2}$, short-dashed line;
$\epsilon_{4}$, dotted line.} \label{fig2}
\end{figure}

\begin{figure}
\caption{The shape of the released pulses depending on the order
of switching on of the control pulses: (a) the pulse (4) switched
on before the pulse (2); (b) the pulse (4) switched on after the
pulse (2). The line styles as in Fig. \ref{fig2}b. }

\label{fig3}
\end{figure}

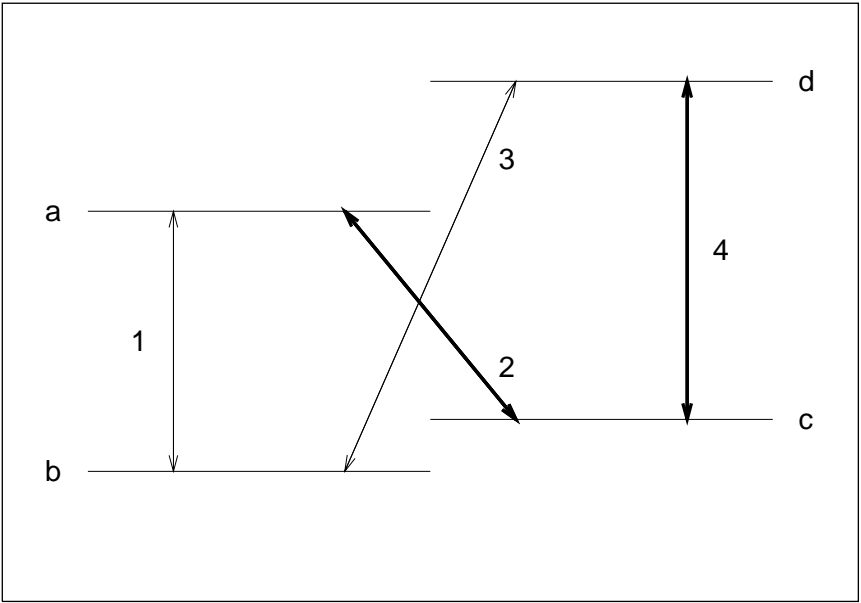
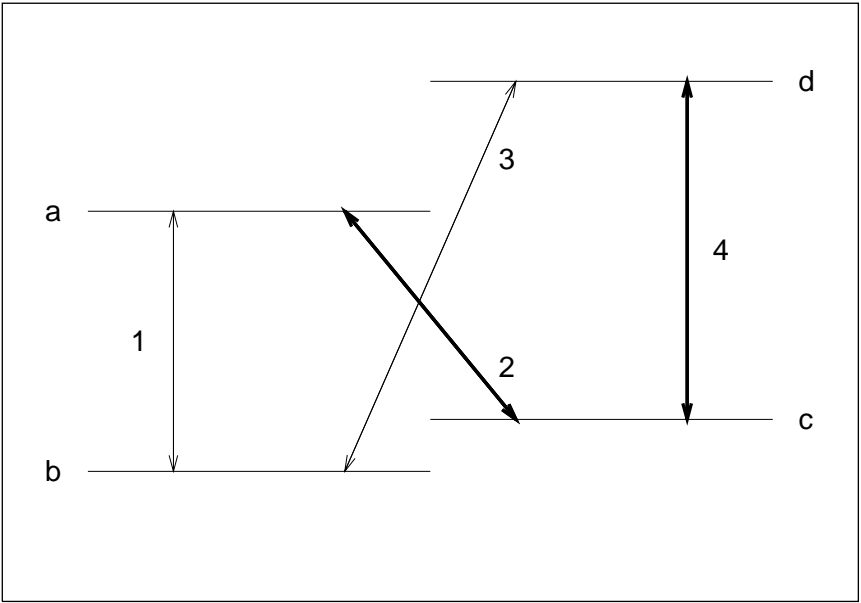
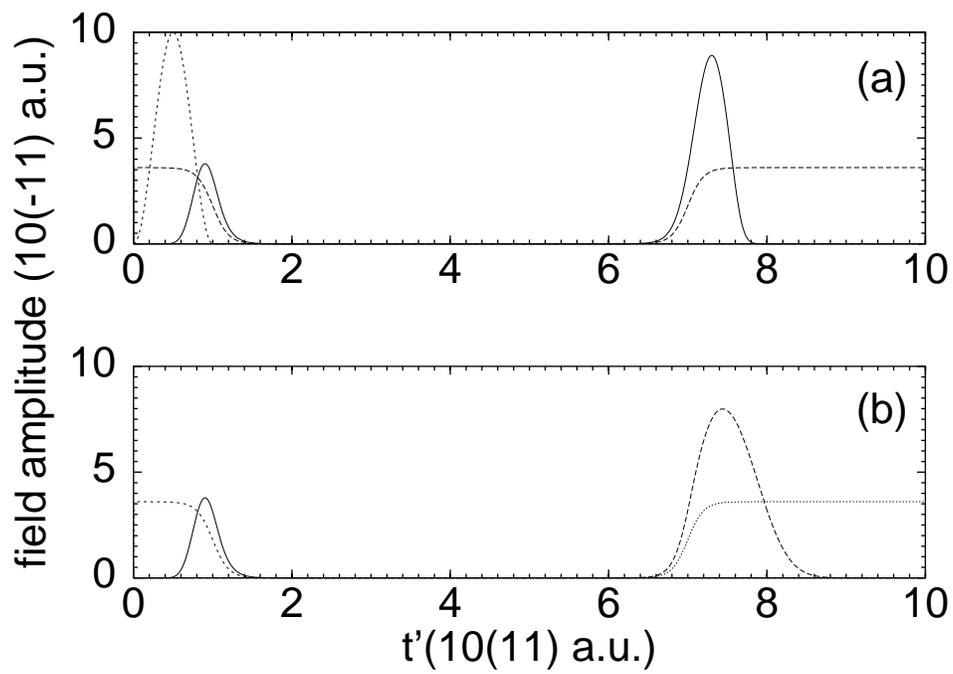
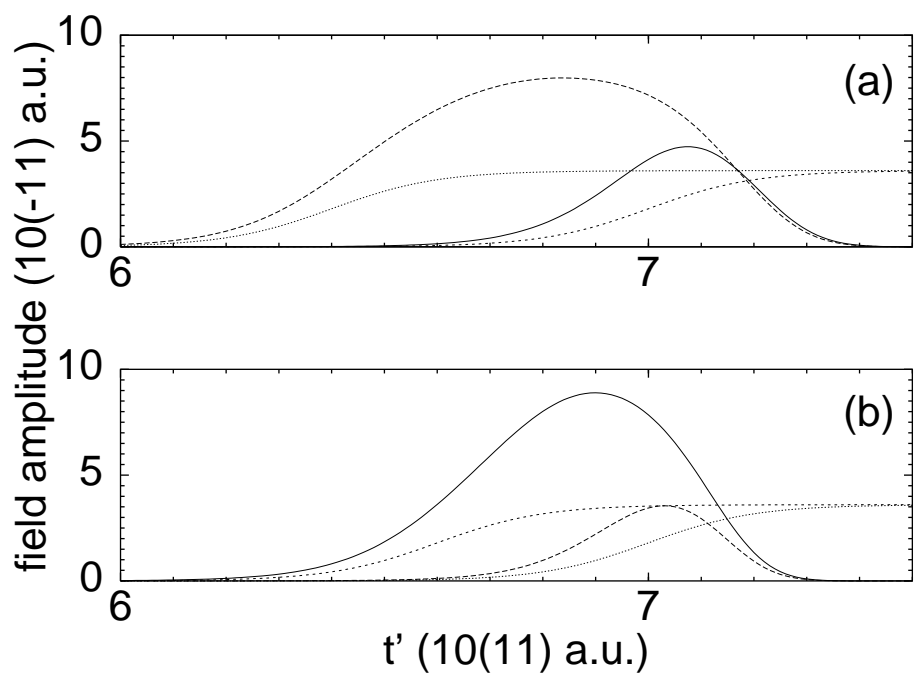
\end{document}